\documentclass[12pt]{article}
\usepackage{amssymb,graphicx}
\usepackage{epstopdf}
\usepackage{amsmath,amsfonts}
\usepackage{epsfig}

\def\d{\partial}

\newcommand{\be}{\begin{equation}}
\newcommand{\ee}{\end{equation}}
\newcommand{\bez}{\begin{equation*}}
\newcommand{\eez}{\end{equation*}}
\newcommand{\bea}{\begin{eqnarray}}
\newcommand{\eea}{\end{eqnarray}}
\newcommand{\bg}{\begin{gather}}
\newcommand{\eg}{\end{gather}}
\newcommand{\bseq}{\begin{subequations}}
\newcommand{\eseq}{\end{subequations}}

\begin{document}
\begin{flushright}
%Prepint Number
\end{flushright}
\vspace{10pt}
\begin{center}
  {\LARGE \bf Harrison--Zeldovich spectrum\\[0.3cm] from conformal
invariance} \\
\vspace{20pt}
%\medskip
V.A.~Rubakov\\
\vspace{15pt}
\textit{
Institute for Nuclear Research of
         the Russian Academy of Sciences,\\  60th October Anniversary
  Prospect, 7a, 117312 Moscow, Russia}\\
\end{center}

\begin{abstract}
We show that flat spectrum of small perturbations of field(s)
is generated in a simple way in a theory of multi-component scalar
field provided this theory is conformally invariant, it has some
global symmetry and the quartic potential is negative. We suggest a
mechanism of converting these field perturbations into adiabatic
scalar perturbations with flat spectrum.
\end{abstract}

\section{Introduction}

An elegant mechanism of the generation of (almost) flat spectrum
of scalar perturbations is provided by inflation~\cite{inflation-HZ}.
Yet it is legitimate to ask  whether  the spectrum of the
Harrison--Zeldovich type can emerge in a different way.
This question is of interest, in particular, 
from the viewpoint of alternatives to
inflation --- pre-Big Bang scenario~\cite{pre-BB-i} 
(for a review see Ref.~\cite{pre-BB}), 
ekpyrotic/cycling 
models~\cite{ekpyro-i,cyclic} (for a review see Ref.~\cite{ekpyro}), 
starting-the-Universe picture~\cite{starting},
etc. One possibility 
is to make use of scalar fields with negative
exponential potentials~\cite{finelli,minus-exp,minus-bis,minus-bis2}
(see also Ref.~\cite{minus-old}).
%, which serves as curvaton.
In this way one indeed obtains almost flat spectrum 
of scalar perturbations
in ekpyrotic
models, with overall homogeneity and isotropy of the Universe
%not 
%necessarily 
%ruined
preserved
during the contracting phase because of the stiff  effective equation 
of state~\cite{smooth}. In concrete models of this type
the contracting solutions are
unstable~\cite{koyama,tolley}, 
so that one has to
fine tune initial data in 
%order that 
our part of
the Universe.
% be almost
%homogeneous and isotropic in the end. 
It has been argued,
however, that such
a fine tuning may be automatic,  in a sense,
in a cyclic scenario
with long periods of dark energy domination~\cite{antro-fine-tune}.

In this paper we suggest another mechanism for  generating
the Harrison--Zeldovich spectrum. The idea is to relate scale invariance
of this spectrum to conformal invariance of 
%the relevant 
field theory.
We will see in Section~\ref{sec-2}
that this can be done by introducing conformal scalar field 
as a source of entropy perturbations, which are converted into
adiabatic perturbations at later epoch. 
The three basic ingredients are that (i) the theory possesses
some global symmetry, (ii) the quartic scalar potential is negative
and (iii) there is long enough stage of the cosmological evolution
prior to the conventional
hot Big Bang epoch. At that early stage, 
the classical scalar field rolls down its
potential along the late time attractor. In this regime, the
vacuum perturbations of field(s)
orthogonal to the radial direction --- 
the direction along which the classical field rolls down --- 
get amplified and freeze out having the
spectrum which is
automatically flat. 
%It is the latter, orthogonal field(s) that will serve
%as curvaton.
Hence, getting flat spectrum of the field perturbations
is
reasonably simple.

%\begin{figure}
%\begin{center}
%\includegraphics[scale=1]{phi4.eps}
%\end{center}
%\caption{}
%\label{fig1}}

\begin{figure}[tb!]
\begin{center}
\includegraphics[width=0.5\textwidth,angle=-90]{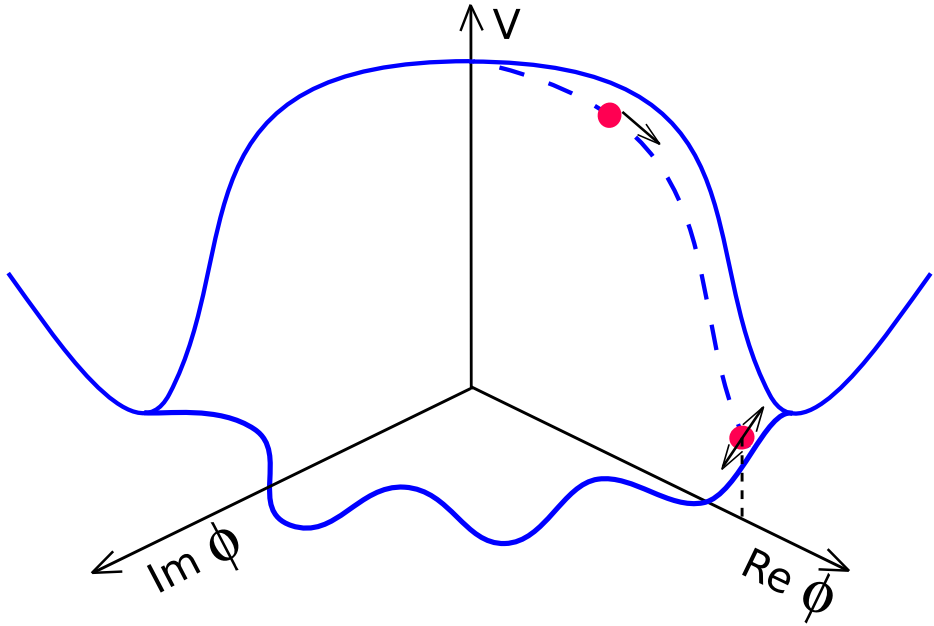}
\end{center}
\caption{The scalar potential is negative quartic and respects global
symmetry ($U(1)$ here) at relatively small $\phi$, and has one or more minima
at large $\phi$. Dashed line shows the evolution of the field
(dot with arrow) at the rolling stage. At that stage the perturbations
(shown by double arrow)
of the pseudo-Goldstone field are developed.
At the hot Big Bang epoch, the pseudo-Goldstone field transfers its energy
to hot matter, and its perturbations are converted into adiabatic 
perturbations. 
\label{fig1}}
\end{figure}

Implementing this mechanism to obtain adiabatic matter perturbations of
correct amplitude is somewhat less 
straightforward. We propose the
corresponding scenario in 
Section~\ref{sec-3}. At the end of the rolling stage, both conformal invariance
and global symmetry are assumed to be broken, and the scalar potential
 is assumed to have a minimum or a set of minima, see Fig.~\ref{fig1}. 
If the scale of global symmetry
breaking is small enough, the situation becomes analogous to the axion 
misalignment picture (for a review see Ref.~\cite{axion-misalign}),  
the analog of axion being
the orthogonal, pseudo-Goldstone field(s) corrseponding to  broken
global symmetry. Unlike in the case of axion, the interactions of the
pseudo-Goldsone field with matter
are not negligible, so this field plays the same role as
curvaton in inflationary models. 
It gets frozen at a slope of the potential, later on
it starts to oscillate and eventually these oscillations convert into
conventional particles. In this way the adiabatic perturbations
with flat spectrum are produced.

%The restrictions on the parameters of our
%model come from the requirements that (i) 
%the scalar perturbations should have the correct value, 
%(ii) the mechanism should work down to the lowest
%length scales observed. 
%In the end of Section~\ref{sec-3} we also discuss the requirement that (iii) 
%the linearized treatment should be valid for all momentum
%scales of interest and for all field perturbations, including the
%radial ones.
%We will see that the latter  requirement, combined with the former two,
%constrains the model quite strongly. 
%In particular, the number of minima of the potential
%for the pseudo-Goldstone field should be very large.
%However, we  argue in 
%Section~\ref{sec-4} that the requirement (iii) may be relaxed.
%If so, the constraints on the
%parameters of our model are not very
%strong (though the multiplicity of the minima of the
%scalar potential is still needed). 

In Section~\ref{sec-4} we discuss a subtle point of our model.
Namely, at the stage when the radial field rolls down, this field
develops its own perturbations. Their growth is fast, and at some point
the linearized theory breaks down. We argue, however, that
the results of the linearized theory concerning the perturbations
of the pseudo-Goldstone field remain valid, and the flat spectrum
of adiabatic perturbations is generated indeed.

We end up in Section~\ref{sec-conclude} with concluding remarks.

\section{Getting flat spectrum}
\label{sec-2}

To begin with, let us consider massless 
scalar field $\phi$ conformally coupled to gravity
and evolving in spatially flat 
4-dimensional FRW Universe. 
Let us assume 
for the time being that there is exact global symmetry,
which we take for definiteneness to be $U(1)$
(though the argument is valid for any compact symmetry group).
Hence, the field is comlex, and the global symmetry is
$\phi \to \mbox{e}^{i\alpha} \phi$.
Conformal invariance allows for quartic scalar potential, which
we assume to be negative, 
\bez
V(\phi) = - h^2 |\phi|^4
\eez
(modulo an additive constant).
As usual, one defines the field $\chi$ by
\bez
\phi = \frac{\chi}{a} \equiv \frac{\chi_1 + i \chi_2}{a} \; .
\eez
Then the action for $\chi$ in conformal coordinates is
\bez
S_\chi = \int~d^3x~d\eta~
\left[
\d_\mu \chi^* \d^\mu \chi + h^2 |\chi|^4
\right] \; ,
\eez
where indices are raised by Minkowski metric.

Let us consider spatially homogeneous background field.
The field equation is
\be
- \chi^{\prime \prime} + 2h^2 |\chi|^2 \chi = 0 \; ,
\label{2}
\ee
where prime denotes $d/d\eta$.
In terms of the radius and phase, $\chi=\rho \mbox{e}^{i\theta}$,
one of the equations is the conservation of current,
\bez
\frac{d}{d\eta} \left( \rho^2 \theta^\prime \right) = 0 \; .
\eez
Hence, as the value of $\rho$ increases, the phase $\theta$ freezes
out, and the evolution proceeds along the radial direction.
 Without 
loss of generality we  take $\chi$ to be real 
at that time, $\chi=\chi_1$. Then the
first integral of the equation of motion is
$ \chi_1^{\prime \, 2} - h^2 \chi_1^4 = \epsilon = \mbox{const}$.
Hence, the solutions approach the asymptotics
that is independent of $\epsilon$,
\be
\chi_1 (\eta) = \frac{1}{h (\eta_* - \eta)} \; ,
\label{16-1}
\ee
where $\eta_*$ is a constant of integration.

Now, let us study field perturbations in the orthogonal direction,
$\delta \chi_2$. They obey the linearized equation, in momentum
representation,
\be
(\delta \chi_2)^{\prime \prime}
+ k^2 \delta \chi_2 - 2 h^2 \chi_1^2 \delta \chi_2  = 0 \; .
\label{1}
\ee
At early times, when $k(\eta_* - \eta) \gg 1$, the second term dominates,
and $\delta \chi_2$ oscillates like free scalar field in Minkowski
space-time. At later times the third term dominates instead. 
So, the perturbation
``exits the horizon'', but now the ``horizon'' is due to evolving
$\chi_1$ (hereafter we use quotation marks to distinguish this
``horizon'' from the true cosmological horizon). 
Outside the ``horizon'', i.e., at $k(\eta_*- \eta) \ll 1$, the perturbation
$\delta \chi_2$ evolves in the same way as the background $\chi_1$.
%,
%\be
%\delta \chi_2 (\eta) = \mbox{const} \cdot \chi_1 (\eta)
%\ee
%where constant depends on conformal momentum $k$. 
This is clear from the fact that at $k=0$, eq.~\eqref{1} is the linearized
equation \eqref{2}. Now, alongside with the solution $\chi_1 (\eta)$,
eq.~\eqref{2} has a solution  $\mbox{e}^{i\alpha}  \chi_1 (\eta)$, 
where $\alpha$ is a real constant. For small $\alpha$ the latter solution
is $(\chi_1 + i\alpha  \chi_1) $, and 
the second part is small perturbation, 
which is precisely $\delta \chi_2$. So, if the perturbation $\delta \chi_2$
oscillates with amplitude $c(k)$ at early times, it behaves at late times as
\be 
\delta \chi_2 = C c(k) \chi_1 (\eta) 
  = C^\prime \frac{c(k)}{k (\eta_* - \eta)} \; , 
\label{3}
\ee
%
%
%\begin{subequations}
%\begin{align}
%\delta \chi_2 & = C c(k) \chi_1 (\eta) 
%\\
% & = C^\prime \frac{c(k)}{k (\eta_* - \eta)}
%\label{3}
%\end{align}
%\end{subequations}
where the factor $k^{-1}$ is evident on dimensional grounds, $C$ and $C^\prime$
are time-independent, and $C^\prime$  
is independent of $k$.
This is precisely the enhancement of perturbations needed for
getting flat spectrum of $\delta \chi_2$.
We stress that this property is entirely due to the symmetries of the
model: the behaviour \eqref{16-1} of the background is a 
consequence of conformal invariance, whereas the behaviour
\eqref{3} of perturbations is a consequence of the global
$U(1)$ symmetry. 

In more detail, eq.~\eqref{1} reads
\bez
(\delta \chi_2)^{\prime \prime}
+ k^2 \delta \chi_2 - \frac{2}{(\eta_* - \eta)^2} \delta \chi_2  = 0 \; .
\eez
This is formally the same equation as the equation for perturbations of
minimally coupled massless scalar field in de~Sitter space-time.
Hence, the spectrum of $\delta \chi_2$ is the same flat spectrum.
We are interested in the solution that behaves at early times  as
\be
\chi_2^{(-)} = \frac{1}{(2\pi)^{3/2}\sqrt{2k}} \mbox{e}^{ik (\eta_* - \eta)}
\; .
\label{1b}
\ee
Modulo overall constant
phase, this solution is expressed through the Hankel function,
\bez
\chi_2^{(-)} = \frac{1}{4\pi} \sqrt{\frac{\eta_* - \eta}{2}}
H_{3/2}^{(1)} \left[k (\eta_* - \eta) \right] \; .
\eez
At $k (\eta_* - \eta) \ll 1$ this solution is
\be
\chi_2^{(-)} = \frac{i}{2\pi^{3/2}} \frac{1}{k^{3/2} (\eta_* - \eta)}
\; , 
\label{1a}
\ee
in accord with \eqref{3}. Assuming that the quantum field $\delta \chi_2$
is originally in the vacuum state, we obtain its power spectrum
outside the ``horizon'' in the standard way. 
%The latter expression determines
%the power spectrum for the quantum field
%\be
%\delta \chi_2 = \int~d^3k~ \left(\chi_2^{(-)}({\bf k}) A_{\bf k} + h.c.
%\right)
%\ee
%One finds
%\be
%\langle (\delta \chi_2)^2 ({\bf x}) \rangle
%= \frac{1}{\pi \Gamma^2 (-1/2)} \frac{1}{(\eta_* - \eta)^2}
%\int~\frac{dk}{k} 
%\label{3a}
%\ee
%This means that 
The spectrum is flat, and its amplitude is
\be
\Delta_{\chi_2} \equiv \sqrt{{\cal P}_{\chi_2}} = \frac{1}{2\pi
(\eta_* - \eta)} \; .
\label{10a}
\ee
%If the field $\chi_2$ is curvaton, it generates flat spectrum
%of scalar perturbations.
As promised, generating flat spectrum of the field perturbations
is fairly straightforward.

%Note that in the case of exact conformal invariance,
%the spectrum is exactly flat. If conformal invariance is
%slightly broken, the spectrum would probably have small tilt.

We end up this section by noting that 
the mechanism requires long time during which
the classical field $\phi$ rolls down its potential. Indeed, 
let $\eta_i$ be the time at which the field starts to roll down, and
$\eta_f$ the time at which rolling down terminates.
Then  
the existence of the
regimes \eqref{1b} and \eqref{1a} requires  that
\begin{eqnarray*}
&& k(\eta_* - \eta_i) \gg 1 \; ,
\\
&& k(\eta_* - \eta_f) \ll 1 \; .
\end{eqnarray*}
These inequalities should hold for all scales of interest, say, for
present momenta (we set the present value of the scale factor equal to 1)
\be
k_{min} \sim (1~\mbox{Gpc})^{-1} \lesssim k \lesssim k_{max} \sim
(100~\mbox{kpc})^{-1} \; .
\label{16-2}
\ee
Therefore, the duration of the rolling stage should be large,
\be
(\eta_f - \eta_i) \gg (k_{min}^{-1} - k_{max}^{-1}) \approx k_{min}^{-1}
\; .
\label{17-1}
\ee
This excludes the possibility that the mechanism works at the
hot stage of the cosmological evolution: there is simply not enough
time. On the other hand, there may well be enough time in the ekpyrotic
or starting-the-Universe scenario\footnote{Note that the inequality
\eqref{17-1} means that the comoving size of light cone originating from
$\eta=\eta_i$ and measured at $\eta=\eta_f$ exceeds the comoving
size of the visible Universe. Hence, the mechanism can work only in
cosmological models which solve, at least formally, the horizon problem.}.

\section{Implementing the mechanism}
\label{sec-3}

In this Section we suggest a
way to implement the  mechanism and generate adiabatic perturbations
from the perturbations $\delta \phi_2$.
Let us assume that as the classical field $\phi_1 = \chi_1/a$ approaches the
region $\phi_1 \sim f$, conformal symmetry gets broken and the  
potential along the radial direction
has a minimum at $\phi_1=f$, where $f$ is some high energy
scale. 
%Let us proceed by keeping $O(2)$ symmetry intact. In fact,
We  assume that $O(2)$ symmetry is broken as well, but that
the scale of
that breaking is small compared to $f$, so we neglect
the effects of  $O(2)$ symmetry breaking for the time being. Then the 
classical field
$\phi_1$ oscillates near $f$ and eventually settles down to $\phi_1 = f$
(say, by producing conventional particles).
We require that this process occurs 
%at ``pre-historic'' stage of the
%cosmological evolution; another possibility is that it occurs in hot 
%expanding Universe but 
at the time when the energy density of the
field $\phi_1$ is much smaller than the total energy density 
in the Universe.
%
%of hot matter.
The reason for this  requirement
is that the field $\phi_1$ develops its own perturbations.
In terms of perturbations $\delta \chi_1$, the linearized equation for this
mode is
\bez
(\delta \chi_1)^{\prime \prime}
+ k^2 \delta \chi_1 - \frac{6}{(\eta_* - \eta)^2} \delta \chi_1  = 0 \; .
\eez
The solution which is negative frequency at early times is
 again expressed through the Hankel function, but with another index,
\bez
\chi_1^{(-)} = \frac{1}{4\pi} \sqrt{\frac{\eta_* - \eta}{2}}
H_{5/2}^{(1)} \left[k (\eta_* - \eta) \right] \; .
\eez
At $k(\eta_* - \eta) \ll 1$ this solution is
\be
\chi_1^{(-)} = - \frac{i}{\pi \Gamma (-3/2)} \frac{1}{k^{5/2} 
(\eta_* - \eta)^2} \; .
\label{2a}
\ee
This behaviour
again has simple interpretation: if $\chi_1$ is a classical solution
to the field equation \eqref{2}, then 
\be 
\delta \chi_1 \propto \chi_1^\prime
\label{6a}
\ee
is a solution 
to the linearized classical field
equation for $k \to 0$. Hence $\delta \chi_1 \propto
(\eta_* - \eta)^2$ on super-''horizon'' scales; the delpendence on $k$
again follows from dimensional argument. In any case, eq.~\eqref{2a}
implies that the spectrum of perturbations of the field
$\chi_1$ is red,
\be
\Delta_{\chi_1} \equiv \sqrt{{\cal P}_{\chi_1}} = \frac{3}{2\pi
k(\eta_* - \eta)^2} \; .
\label{5a}
\ee
These perturbations should not leave any trace in the late Universe; 
%the usual matter; 
the simplest way to achieve that is to impose the above
requirement.

The perturbations of the field $\phi_1$ may become so strong that
they ruin the linearized analysis. 
To be on the safe side, we impose here the 
requirement that the field $\phi_1$ is approximately homogeneous
over the whole visible Universe at the time when the scales of interest
exit the ``horizon'' (see, however, Section~\ref{sec-4}). 
This is sufficient for 
the linearized theory to be valid 
in our patch of the Universe at that time; 
perturbations $\delta \phi_1$ 
of longer wavelengths correspond to unobservable
shift of the parameter $\eta_*$ in the whole visible Universe,
see \eqref{6a}. 
Our requirement gives
\be
\Delta_{\chi_1} (k_{min}, \eta_{max}) \ll \chi_1 (\eta_{max}) \; ,
\label{19-1}
\ee
where $k_{min}^{-1}$ is the comoving size of the visible Universe,
cf. \eqref{16-2}, and $\eta_{max}$ is the time of the 
``horizon'' exit of the shortest interesting modes. The value of 
$\eta_{max}$ is determined 
by
\bez
k_{max}(\eta_* - \eta_{max}) \sim 1 \; .
\eez
Making use of \eqref{16-1} and \eqref{5a} we obtain a constraint
on the coupling constant 
\be
 h \ll \frac{k_{min}}{k_{max}} \sim 10^{-4} \; .
\label{11a}
\ee
We will further discuss the non-linearity issue in Section~\ref{sec-4}.

Let us now turn to the field $\phi_2$. In fact, after the field
$\phi_1$ settles down to $f$, it is more appropriate to
speak about the phase $\theta$
of the original field $\phi$. We keep the notation $\phi_2$,
with understanding that $\theta = \phi_2/f$.

Let us first estimate the amplitude of perturbations
$\delta \phi_2$ at the time $\eta_f$ when the background field
$\phi_1$ reaches $f$. At that time one has
\be
\chi_1 (\eta_f) \sim \frac{1}{h (\eta_* - \eta_f)} \sim a_f \cdot f
\; ,
\label{8a}
\ee
where the subscript $f$ refers to the time $\eta_f$. From 
\eqref{10a} we obtain
\bez
\Delta_{\phi_2} (\eta_f) = \frac{\Delta_{\chi_2}(\eta_f)}{a_f}
\simeq \frac{1}{2\pi} hf \; .
\eez
We now have to convert these entropy perturbations into adiabatic ones.

Let us recall that
the global $O(2)$ symmetry is broken near $|\phi| = f$.
We have in mind axion-like picture, in which the energy scale of this
breaking is smaller than $f$. Importantly, the phase
field $\phi_2$, as any  (pseudo-)Goldstone field, {\it minimally}
couples to gravity~\cite{voloshin}.
From this point on, the discussion is parallel to that in the
curvaton model.

Generically, $\phi_2=0$ is not a minimum of the scalar potential,
so the radial evolution ends up at a slope of the potential
for the pseudo-Goldstone field $\phi_2$. 
Let us assume that at the time $\eta_f$, 
the Hubble parameter exceeds the mass parameter of this
pseudo-Goldstone field.
In this case the field $\phi_2$
stays close to zero for some time\footnote{This is true both for
expanding Universe and for contracting Universe dominated by matter
with stiff equation of state, $w>1$.}, 
then rolls down to the nearest minimum
of its potential,
oscillates there and eventually the oscillations decay into usual particles.
We assume that the decay happens at the hot Big Bang 
stage of the cosmological 
expansion, when the Universe is either radiation-dominated, or
temporarily dominated by the oscillating field $\phi_2$ itself.
Let $M$ denote the distance to the nearest minimum of the 
potential for the filed $\phi_2$, so that the number of minima
of the scalar potential is $N \sim f/M$. Then at the beginning
of oscillations and until their decay, 
the perturbations in the energy density of the pseudo-Goldstone field are
\bez
\frac{\delta \rho_{\phi_2}}{\rho_{\phi_2}} \sim
\frac{\delta \phi_2 (\eta_f)}{M} \; .
%\sim 
%\frac{\delta \chi_2 (\eta_f)}{a_f M} 
\eez
This leads to adiabatic perturbations with flat spectrum
and the amplitude
\be
\Delta \sim r \frac{\Delta_{\chi_2} (\eta_f)}{a_f M}
\sim r \frac{hf}{2\pi M} \; ,
\label{12a}
\ee
where $r$ is the ratio of the energy density of the field $\phi_2$
to the total energy density at the time the oscillations decay.

The pseudo-Goldstone picture with a single minimum of the potential for 
the phase would correspond to $M \sim f$. If we insist on
the constraint
\eqref{11a}, the amplitude \eqref{12a}
of the resulting adiabatic perturbations would be too small in that case
(barring fine tuning). On the other hand, with $N \sim f/M \gg 1$ the required
amplitude $\Delta \sim 10^{-5}$ can obtained without fine tuning of 
other parameters\footnote{This 
discussion assumes that after the rolling stage,
the value of the phase of the field $\phi$
is generic. Alternatively, one could make use of the anthropic argument
for explaining why the field $\phi_2$ ends up close to the minimum of 
its potential, i.e., why $M \ll f$, so that the amplitude of the matter
perturbations is large enough,
cf. Refs.~\cite{tegmark,wilczek}.
With this line of reasoning, having large  number of minima of the potential
for the pseudo-Goldstone field is certainly 
unnecessary.}. 
As we discuss in the end of Section~\ref{sec-4}, the constraint
\eqref{19-1}, and hence \eqref{11a} may in fact be unnecessary, so
the mechanism may work for $N=1$ as well.

\section{Generalizing to non-linear radial field perturbations}
\label{sec-4}

Let us come back to the non-linearity issue.
Since the perturbations \eqref{5a} increase in time faster than
the background field \eqref{16-1}, the evolution in the radial direction
may become non-linear at some stage. Naively, one would think that the
analysis of Section~\ref{sec-2} goes through only if
this does not happen, i.e., if the perturbations $\delta \phi_1$ remain
small up to the time $\eta_f$. Such a requirement would push the parameters
of the model to a very contrived range:  the coupling contant $h$ would
have to be extremely small, the field value $f$ would have to be
super-Planckian, and the number of minima of the scalar 
potential, $N \sim f/M$, 
would have to be very large. This is discussed 
in Appendix. However, we argue that non-linearity of the
radial field perturbations
does not spoil the result \eqref{12a}.

The argument is as follows. Once the inequality \eqref{19-1} is
satisfied, all interesting scales exit the ``horizon'' when
the linearized theory is still valid. Let $\eta_{lin}$ be some moment
of time when the linearized theory is valid, but all interesting
scales are already
super-''horizon''. At that time, $\delta \phi_2$ may be considered
as classical random field whose spatial gradients are negigible.
According to \eqref{10a} and \eqref{16-1} this field is small,
$\delta \phi_2 / \phi_1 \sim h$, and it evolves in the same way as
$\phi_1$,
\be
\delta \phi_2 (\eta) = \frac{\delta \phi_2 (\eta_{lin})}{\phi_1 (\eta_{lin})}
\phi_1 (\eta) \; .
\label{19*}
\ee
The field $\phi_1 ({\bf x}, \eta) = \phi_1 (\eta) + \delta \phi_1 
({\bf x}, \eta)$ is also  classical random field which is homogeneous
inside the ``horizon''. At late times this field becomes inhomogeneous,
but
on super-horizon scales only. 
The point is that the evolution of the
entire system proceeds according to the homogeneous classical field
equation. Therefore, the symmetry
argument presented before eq.~\eqref{3} tells 
that $\delta \phi_2$ is still given by \eqref{19*}, but now with
$\phi_1 = \phi_1 ({\bf x}, \eta)$. Note that $\delta \phi_2$ remains small.
The field $\phi_1$ reaches the value $f$
in the end, and at that time
\bez
\delta \phi_2  = \frac{\delta \phi_2 (\eta_{lin})}{\phi_1 (\eta_{lin})}
f \; .
%\label{19*}
\eez
This coincides with the result of the linearized theory.

Another way to phrase this argument is to write the rolling field $\phi_1$,
with long-ranged perturbations included, as
\be
\phi_1 ({\bf x}, \eta) = \frac{1}{h a(\eta) [\eta_{*\; eff} ({\bf x}) - \eta]}
\; .
\label{19-3}
\ee
Comparing this expression with \eqref{5a} at the time when the linearized
theory is still valid, we find that $\delta \eta_*({\bf x})
= \eta_{* \; eff} ({\bf x}, \eta) - \eta_*$ is long ranged 
random classical field
whose fluctuation is
\bez
\Delta_{\delta \eta_*} = \frac{3h}{2\pi k} \; .
\eez
At late times, when $\delta \eta_* \gtrsim (\eta_* - \eta)$,
the dependence of $\phi_1 ({\bf x}, \eta)$ on $\delta \eta_*$ becomes 
non-linear, but the expression \eqref{19-3} remains valid.
The point is, again, that  the expression
\eqref{19*}  with
$\phi_1 = \phi_1 ({\bf x}, \eta)$
is a solution to the linearized field equation for $\delta \phi_2$
at all times. 
To see this explicitly, let us show 
that $|{\boldsymbol \nabla} \delta \phi_2| \ll \delta \phi_2^\prime$,
so that spatial variation of $\eta_{* \; eff}$ can be neglected in
that field equation, and the argument before eq.~\eqref{3}
indeed applies. Using  \eqref{19*} we write
\be
\frac{|{\boldsymbol \nabla} \delta \phi_2|}{\delta \phi_2^\prime}
= |{\boldsymbol \nabla} \delta \eta_*| \; .
\label{19-2}
\ee
Now, ${\boldsymbol \nabla}\delta \eta_*$ is random field with fluctuation
\bez
\Delta_{({\boldsymbol \nabla} \delta \eta_*)} = k \Delta_{\delta \eta_*} 
\sim h \; ,
\eez
so the ratio \eqref{19-2} is indeed small. Since the final state of the
super-''horizon''
perturbation $\delta \phi_2$ is independent of the parameter $\eta_*$,
we see that non-linearity of the evolution of $\phi_1$, encoded in
the non-linear dependence of $\phi_1$ on $\delta \eta_*$,
is irrelevant
for the spectrum of $\delta \phi_2$.

The latter way of presenting the argument suggests that 
the condition \eqref{19-1} is in fact unnecessary. The only requirement
that remains is that $h \ll 1$. This ensures 
that the perturbations  $\delta \phi_1$
become non-linear long after they exit the ``horizon'', and the spatial
variation of $\eta_{* \; eff}$ is irrelevant in the field equation for 
$\delta \phi_2$. Hence, the parameters of our model are constrained very
weakly; in particular, there may exist just one minimum of the
potential for the pseudo-Goldstone field, i.e., $M \sim f$.

\section{Conclusions}
\label{sec-conclude}

We conclude this paper with a few remarks. First, the mechanism for
converting the perturbations of $\phi_2$ into adiabatic
perturbations, suggested in Section~\ref{sec-2}, is most probably not 
unique. For example, one 
can imagine adding other fields which interact with
the pseudo-Goldstone field $\phi_2$ so that the energy density
of these fields,
after $\phi_1$ relaxes to $f$, is linear in $\delta \phi_2$.
In that case the simple formula \eqref{12a} would not be valid.

Second, the resulting spectrum needs not be exactly flat.
One expects that for conformal symmetry  slightly broken during
the stage of rolling $\phi_1$, the spectrum becomes tilted.
The tilt may be present also if the global symmetry ($U(1)$ in our
example) is explicitly broken at that stage. How strong is the
variation of the tilt with $k$, and whether there are other
properties of the spectrum in these cases, remains to be understood.
 
Third, if the coupling constant $h$ is not very small, one would expect
sizeable non-Gaussianity of the perturbations. Also, there may be
an  interplay between the perturbations $\delta \phi_1$ and 
$\delta \phi_2$ which may result in peculiar properties
of the adiabatic perturbations. It would be interesting to see
whether or not the latter may help discriminate our
mechanism from others (say, from inflationary ones). 

Fourth, as in many other cases, the generation of scalar
perturbations in our model is unrelated to the generation of tensor
modes. Thus, there is no reason to expect sizeable gravitational wave
background in our scenario.

Finally, we left aside the issue of the initial conditions for the
field $\phi$. We simply assumed that this field is spatially
homogeneous and starts rolling from near the top of its quartic 
potential. Whether these initial conditions may emerge naturally
in the context of some cosmological scenario remains unclear.

We hope to address these and other issues in future.

The author is indebted to D.~Gorbunov, A.~Khmelnitsky, E.~Nugaev,
G.~Rubtsov and I.~Tkachev for helpful discussions. This work
has been supported in part by Russian Foundation for Basic
Research grant 08-02-00473.

\section*{Appendix}

In this Appendix we discuss what kind of constraints
on the parameters one would obtain by
insisting that the entire evoluton of both perturbations
$\delta \phi_1$ and $\delta \phi_2$
occurs in the linear regime, so that the analysis of Section~\ref{sec-2}
is literally valid. The strongest constraint comes from the
requirement that the perturbations of the radial field $\delta \phi_1$
are small by the end of the regime of rolling $\phi_1$,
\be
\delta \phi_1 (\eta_f) \sim \Delta_{\chi_{1}} (\eta_f)/a_f \ll f \; .
\label{4a}
\ee
Let us pretend that this inequality should be valid 
for length scales up to the present 
horizon size $k_{min}^{-1}$;
perturbations of longer wavelengths correspond to unobservable
shift of the parameter $\eta_*$ in the whole visible Universe. 

Making use of \eqref{8a} and \eqref{5a} one finds that the condition
\eqref{4a} gives
\be
   \frac{k_{min}}{a_f} \gg h^2 f \; .
\label{7a}
\ee
Given the very small value of $k_{min}$ of the order of
the present Hubble parameter, the latter inequality
impies that the coupling constant $h$ is extremely small.
Indeed, assuming that the Universe did not expand significantly
before the beginning of the conventional hot Big Bang stage
(no inflation) and ignoring weak dependence on the effective number of
degrees of freedom $g_*$, one finds that $a_f \gtrsim T_0/T_{max}$,
where $T_{max}$ and $T_0$ are the maximum temperature in the
Universe and the present temperature, respectively.
Hence, the constraint \eqref{7a} reads
\be
h^2 f \ll \frac{k_{min}}{T_0} T_{max} \sim 10^{-29} T_{max} < 10^{-29} M_{Pl}
\; .
\label{2*}
\ee
The constraint \eqref{7a}  implies also that 
\be
   f \gg M_{Pl} \; .
\label{2+}
\ee
This comes out from the estimate of the temperature $T_{end}$ at which the
oscillations of the pseudo-Goldstone field $\phi_2$ are converted into 
usual particles. At this temperature the perturbations $\delta \phi_2$
are transformed
into adiabatic ones, and one requires that $T_{end} \gtrsim 100$~GeV
in order  that the baryon asymmetry and dark matter may be thermally
produced afterwards. To estimate $T_{end}$ in our model, we notice that
the oscillations of $\phi_2$ begin at the time $t_{osc}$ when
$H(t_{osc}) \sim \mu$ where $\mu^2$ is the curvature of the potential
for the pseudo-Golstone
field. Again ignoring the dependence on $g_*$ we obtain the energy densities
of hot matter and the pseudo-Goldstone field at that time,
\bez
\rho_{rad} (t_{osc}) \sim T^4 (t_{osc}) \sim \mu^2 M_{Pl}^2 \; , 
\;\;\;\;\;\;\; \rho_{\phi_2} (t_{osc}) \sim \mu^2 M^2 \; .
\eez
The oscillations end up when $\rho_{rad}/\rho_{\phi_2} = r \leq 1$,
so one finds
\bez
\frac{T_{end}}{T_{osc}} \sim \frac{a(t_{osc})}{a(t_{end})}
= \frac{1-r}{r} \frac{\rho_{\phi_2} (t_{osc})}{\rho_{rad}(t_{osc})} \; .
\eez
Hence
\bez
T_{end} \sim \frac{(1-r) M^2 }{r M_{Pl}^2} \sqrt{\mu M_{Pl}} \; .
\eez
Making use of \eqref{12a} we find
\bez
T_{end} \sim \frac{h^2 f^2 r (1-r)}{\Delta^2 M_{Pl}^2} \sqrt{\mu M_{Pl}}
\; .
\eez
Finally, using \eqref{2*} and inserting $\Delta \sim 10^{-5}$
we obtain
\bez
T_{end} \ll \left(\frac{f \mu^{1/2}}{M_{Pl}^{3/2}} \right)\; \mbox{GeV}
\; .
\eez
Since $\mu^2$ is the curvature of the scalar potential, one has 
$\mu < M_{Pl}$. Therefore, high enough value $T_{end} \gtrsim 100$~GeV
can be obtained only for super-Planckian values of $f$. Together
with \eqref{2*} this implies that the coupling constant is extremely
small. In turn, the smallness of $h$ and eq.~\eqref{12a}
mean that the correct amplitude of
perturbations  can be obtained only for huge number of
minima of the scalar potential, $N=M/F$.


\begin{thebibliography}{99}
\bibitem{inflation-HZ}
V.~F.~Mukhanov and G.~V.~Chibisov,
  %``Quantum Fluctuation And Nonsingular Universe. (In Russian),''
  JETP Lett.\  {\bf 33} (1981) 532
  [Pisma Zh.\ Eksp.\ Teor.\ Fiz.\  {\bf 33} (1981) 549];\\
  %%CITATION = ZFPRA,33,549;%%
S.~W.~Hawking,
  %``The Development Of Irregularities In A Single Bubble Inflationary
  %Universe,''
  Phys.\ Lett.\  B {\bf 115} (1982) 295;\\
  %%CITATION = PHLTA,B115,295;%%
A.~A.~Starobinsky,
  %``Dynamics Of Phase Transition In The New Inflationary Universe Scenario And
  %Generation Of Perturbations,''
  Phys.\ Lett.\  B {\bf 117} (1982) 175;\\
  %%CITATION = PHLTA,B117,175;%%
A.~H.~Guth and S.~Y.~Pi,
  %``Fluctuations In The New Inflationary Universe,''
  Phys.\ Rev.\ Lett.\  {\bf 49} (1982) 1110;\\
  %%CITATION = PRLTA,49,1110;%%
J.~M.~Bardeen, P.~J.~Steinhardt and M.~S.~Turner,
  %``Spontaneous Creation Of Almost Scale - Free Density Perturbations In An
  %Inflationary Universe,''
  Phys.\ Rev.\  D {\bf 28} (1983) 679.
  %%CITATION = PHRVA,D28,679;%%

\bibitem{pre-BB-i}
G.~Veneziano,
  %``Scale Factor Duality For Classical And Quantum Strings,''
  Phys.\ Lett.\  B {\bf 265} (1991) 287;\\
  %%CITATION = PHLTA,B265,287;%%
M.~Gasperini and G.~Veneziano,
  %``Inflation, deflation, and frame independence in string cosmology,''
  Mod.\ Phys.\ Lett.\  A {\bf 8} (1993) 3701
  [arXiv:hep-th/9309023];\\
  %%CITATION = MPLAE,A8,3701;%%
  M.~Gasperini and G.~Veneziano,
  %``Pre - big bang in string cosmology,''
  Astropart.\ Phys.\  {\bf 1} (1993) 317
  [arXiv:hep-th/9211021].
  %%CITATION = APHYE,1,317;%%

\bibitem{pre-BB}
M.~Gasperini and G.~Veneziano,
  %``The pre-big bang scenario in string cosmology,''
  Phys.\ Rept.\  {\bf 373} (2003) 1
  [arXiv:hep-th/0207130].
  %%CITATION = PRPLC,373,1;%%

\bibitem{ekpyro-i}
  J.~Khoury, B.~A.~Ovrut, P.~J.~Steinhardt and N.~Turok,
  %``The ekpyrotic universe: Colliding branes and the origin of the hot big
  %bang,''
  Phys.\ Rev.\  D {\bf 64} (2001) 123522
  [arXiv:hep-th/0103239];\\
  %%CITATION = PHRVA,D64,123522;%%
J.~Khoury, B.~A.~Ovrut, N.~Seiberg, P.~J.~Steinhardt and N.~Turok,
  %``From big crunch to big bang,''
  Phys.\ Rev.\  D {\bf 65} (2002) 086007
  [arXiv:hep-th/0108187].
  %%CITATION = PHRVA,D65,086007;%%

\bibitem{cyclic}
P.~J.~Steinhardt and N.~Turok,
  %``Cosmic evolution in a cyclic universe,''
  Phys.\ Rev.\  D {\bf 65} (2002) 126003
  [arXiv:hep-th/0111098];\\
  %%CITATION = PHRVA,D65,126003;%%
J.~Khoury, P.~J.~Steinhardt and N.~Turok,
  %``Designing Cyclic Universe Models,''
  Phys.\ Rev.\ Lett.\  {\bf 92} (2004) 031302
  [arXiv:hep-th/0307132];\\
  %%CITATION = PRLTA,92,031302;%%
P.~J.~Steinhardt and N.~Turok,
  %``The cyclic model simplified,''
  New Astron.\ Rev.\  {\bf 49} (2005) 43
  [arXiv:astro-ph/0404480].
  %%CITATION = ASTRE,49,43;%%


\bibitem{ekpyro}
J.~L.~Lehners,
  %``Ekpyrotic and Cyclic Cosmology,''
  Phys.\ Rept.\  {\bf 465} (2008) 223
  [arXiv:0806.1245 [astro-ph]].
  %%CITATION = PRPLC,465,223;%%

\bibitem{starting}
P.~Creminelli, M.~A.~Luty, A.~Nicolis and L.~Senatore,
  %``Starting the universe: Stable violation of the null energy condition and
  %non-standard cosmologies,''
  JHEP {\bf 0612} (2006) 080
  [arXiv:hep-th/0606090].
  %%CITATION = JHEPA,0612,080;%%

\bibitem{finelli}
 F.~Finelli,
  %``Assisted contraction,''
  Phys.\ Lett.\  B {\bf 545} (2002) 1
  [arXiv:hep-th/0206112].
  %%CITATION = PHLTA,B545,1;%%

\bibitem{minus-exp}
J.~L.~Lehners, P.~McFadden, N.~Turok and P.~J.~Steinhardt,
  %``Generating ekpyrotic curvature perturbations before the big bang,''
  Phys.\ Rev.\  D {\bf 76} (2007) 103501
  [arXiv:hep-th/0702153].
  %%CITATION = PHRVA,D76,103501;%%

\bibitem{minus-bis}
E.~I.~Buchbinder, J.~Khoury and B.~A.~Ovrut,
  %``New Ekpyrotic Cosmology,''
  Phys.\ Rev.\  D {\bf 76} (2007) 123503
  [arXiv:hep-th/0702154].
  %%CITATION = PHRVA,D76,123503;%%

\bibitem{minus-bis2}
P.~Creminelli and L.~Senatore,
  %``A smooth bouncing cosmology with scale invariant spectrum,''
  JCAP {\bf 0711} (2007) 010
  [arXiv:hep-th/0702165].
  %%CITATION = JCAPA,0711,010;%%


\bibitem{minus-old}
%\bibitem{Notari:2002yc}
  A.~Notari and A.~Riotto,
  %``Isocurvature perturbations in the ekpyrotic universe,''
  Nucl.\ Phys.\  B {\bf 644} (2002) 371
  [arXiv:hep-th/0205019];\\
  %%CITATION = NUPHA,B644,371;%%
%
%\bibitem{DiMarco:2002eb}
  F.~Di Marco, F.~Finelli and R.~Brandenberger,
  %``Adiabatic and Isocurvature Perturbations for Multifield Generalized
  %Einstein Models,''
  Phys.\ Rev.\  D {\bf 67} (2003) 063512
  [arXiv:astro-ph/0211276].
  %%CITATION = PHRVA,D67,063512;%%

\bibitem{smooth}
  J.~K.~Erickson, D.~H.~Wesley, P.~J.~Steinhardt and N.~Turok,
  %``Kasner and mixmaster behavior in universes with equation of state w >=
  %1,''
  Phys.\ Rev.\  D {\bf 69} (2004) 063514
  [arXiv:hep-th/0312009];\\
  %%CITATION = PHRVA,D69,063514;%%
D.~Garfinkle, W.~C.~Lim, F.~Pretorius and P.~J.~Steinhardt,
  %``Evolution to a smooth universe in an ekpyrotic contracting phase with w >
  %1,''
  Phys.\ Rev.\  D {\bf 78} (2008) 083537
  [arXiv:0808.0542 [hep-th]].
  %%CITATION = PHRVA,D78,083537;%%

\bibitem{koyama}
  K.~Koyama and D.~Wands,
  %``Ekpyrotic collapse with multiple fields,''
  JCAP {\bf 0704} (2007) 008
  [arXiv:hep-th/0703040].
  %%CITATION = JCAPA,0704,008;%%

\bibitem{tolley}
A.~J.~Tolley and D.~H.~Wesley,
  %``Scale-invariance in expanding and contracting universes from two-field
  %models,''
  JCAP {\bf 0705} (2007) 006
  [arXiv:hep-th/0703101].
  %%CITATION = JCAPA,0705,006;%%


\bibitem{antro-fine-tune}
J.~L.~Lehners and P.~J.~Steinhardt,
  ``Dark Energy and the Return of the Phoenix Universe,''
  arXiv:0812.3388 [hep-th].
  %%CITATION = ARXIV:0812.3388;%%

\bibitem{axion-misalign}
 P.~Sikivie,
  %``Axion cosmology,''
  Lect.\ Notes Phys.\  {\bf 741} (2008) 19
  [arXiv:astro-ph/0610440].
  %%CITATION = LNPHA,741,19;%%

\bibitem{voloshin}
M.~B.~Voloshin and A.~D.~Dolgov,
  %``On Gravitational Interaction Of The Goldstone Bosons,''
  Sov.\ J.\ Nucl.\ Phys.\  {\bf 35} (1982) 120
  [Yad.\ Fiz.\  {\bf 35} (1982) 213].
  %%CITATION = YAFIA,35,213;%%

\bibitem{tegmark}
M.~Tegmark and M.~J.~Rees,
  %``Why is the CMB fluctuation level 10^{-5}?,''
  Astrophys.\ J.\  {\bf 499} (1998) 526
  [arXiv:astro-ph/9709058].
  %%CITATION = ASJOA,499,526;%%

\bibitem{wilczek}
A.~D.~Linde,
  %``Inflation And Axion Cosmology,''
  Phys.\ Lett.\  B {\bf 201} (1988) 437;\\
  %%CITATION = PHLTA,B201,437;%%
F.~Wilczek,
  ``A model of anthropic reasoning, addressing the dark to ordinary matter
  coincidence,''
  arXiv:hep-ph/0408167.
  %%CITATION = HEP-PH/0408167;%%

\end{thebibliography}
\end{document}